\documentclass[conference]{sig-alt}

\usepackage[latin1]{inputenc}
\usepackage[T1]{fontenc}
\usepackage{fancyvrb}
\usepackage{times}
\usepackage{amsmath,amssymb}
\usepackage{euler}
\usepackage{url}
\usepackage{graphicx}
\usepackage{framed}
\usepackage[usenames,dvipsnames]{xcolor}
\usepackage{listings}
\usepackage{enumitem}
\usepackage{url}

\setlist{nolistsep}

\lstdefinelanguage{C++1y}{alsolanguage=C++,
                          escapechar=@,
                          breakatwhitespace=true,
                          morekeywords = {alignof, 
                                          decltype, 
                                          concept, 
                                          axiom, 
                                          requires, 
                                          constexpr}}

\lstdefinelanguage{Output}{}

\newcommand{\code}[1]{\lstinline @#1@}

\lstnewenvironment{program}{\lstset{language=C++1y}}{}

\lstnewenvironment{progout}{\lstset{language=Output}}{}

\newcommand{\secref}[1]{$\S$~\ref{#1}}

\begin{document}

\title{tinyNBI: Distilling an API from essential OpenFlow abstractions}

\numberofauthors{3}
\author{
\alignauthor
C. Jasson Casey\\
  \affaddr{Texas A\&M University}\\
\email{jasson.casey@tamu.edu}
\alignauthor
Andrew Sutton\\
  \affaddr{University of Akron}\\
  \email{asutton@uakron.edu}
\alignauthor Alex Sprinston\\
  \affaddr{Texas A\&M University}\\
  \email{spalex@tamu.edu}
}

\maketitle

\begin{abstract}
If simplicity is a key strategy for success as a network protocol OpenFlow is 
not winning. At its core OpenFlow presents a simple idea, which is a network
switch data plane abstraction along with a control protocol for manipulating
that abstraction. The result of this idea has been far from simple: a 
new version released each year, five active versions, complex feature 
dependencies, unstable version negotiation, lack of state machine definition,
etc. This complexity represents roadblocks for network, software, and hardware
engineers. 


We have distilled the core abstractions present in 5 existing versions of
OpenFlow and refactored them into a simple API called tinyNBI. Our work does not
provide high-level network abstractions (address pools, VPN maps, etc.), instead
it focuses on providing a clean low level interface that supports the 
development of these higher layer abstractions. The goal of tinyNBI is to allow 
configuration of all existing OpenFlow abstractions without having to deal with 
the unique personalities of each version of OpenFlow or their level of support 
in target switches.
\end{abstract}

\section{Introduction}
\label{sec:intro}

Software Defined Networking provides the promise of allowing applications to 
control underlying network services without having to know the details of 
specific network equipment. Unfortunately, the OpenFlow protocol does not
truly deliver on that promise. While OpenFlow does provide an interface that
allows software control of switches, it also moves the burden of managing
all that variability up to the programmer of OpenFlow applications. For example,
in OpenFlow 1.3, in order to install an entry in a flow table, the programmer 
must ensure the following for each switch target:
\begin{enumerate}
  \item the target table exists,
  \item the target table has capacity,
  \item the desired matching set is supported,
  \item the desired instruction set is supported,
  \item the desired action set is supported, and
  \item target ports, groups, and meters exist.
\end{enumerate}

It also gets worse. The desired abstractions and semantics, based on an OpenFlow
1.3 model, may not be present on all the target switches. An application has to
handle these variations with abstractions present in some switches and missing 
in others. Several versions of the same OpenFlow application may need to be 
defined in order to accommodate the different versions of switches in the 
network, and each application would need to be cognizant of the variability 
present in each switch.

For network engineers tasked with the development of OpenFlow applications, the
current model is untenable. While the OpenFlow protocols do enable the software-
based definition of network applications, they do so by shifting the burden of
managing variability to the programmer. Managing this degree of variability in
source code is costly; it results in brittle abstractions that are prone to
errors and are difficult to maintain. What is needed is a  programming interface
that, minimally, satisfies the following requirements.
\begin{enumerate}
\item The interface \emph{must} expose clearly defined abstractions that 
      represent SDN dataplane primitives.

\item These abstractions \emph{must} represent structure and features common
      to all versions of the OpenFlow protocol, and structure and semantics
      in a lower version of the protocol must be \emph{a subset} of
      structure and semantics in succeeding versions.

\item The API (Application Programming Interface) \emph{must} provide access to
      these abstractions and the major functionality associated with them,
      including querying capabilities, configuration, and statistics, and
      modifying configuration.


\item The API \emph{must} provide a mechanism for the application to declare
      the set of primitives and capabilities required for execution on a
      connected switch.

\item The API \emph{should} allow for the developer indicate which capabilities
      may be offloaded from a target switch.

\end{enumerate}
An API satisfying these requirements can be used to build OpenFlow applications,
high performance controllers, and application libraries. In this paper, we 
describe TinyNBI, a minimal Northbound Interface (NBI) that satisfies these 
requirements. We present the interface as, first an abstract specification of a 
switch, enumerating its primitive abstractions: connection, datapath, flow 
table, flow, match, instruction, action, port, group, queue, and meter.  
Second, we present a C API for interfacing the primitives of the 
abstract model. The interface is comprised of a small set of C functions that 
can be used to access the major functionality required by the OpenFlow 
specifications. We explain how the interface satisfies the requirements listed 
above.

\begin{figure}[h]
  \begin{center}
    \includegraphics[width=.5\textwidth]{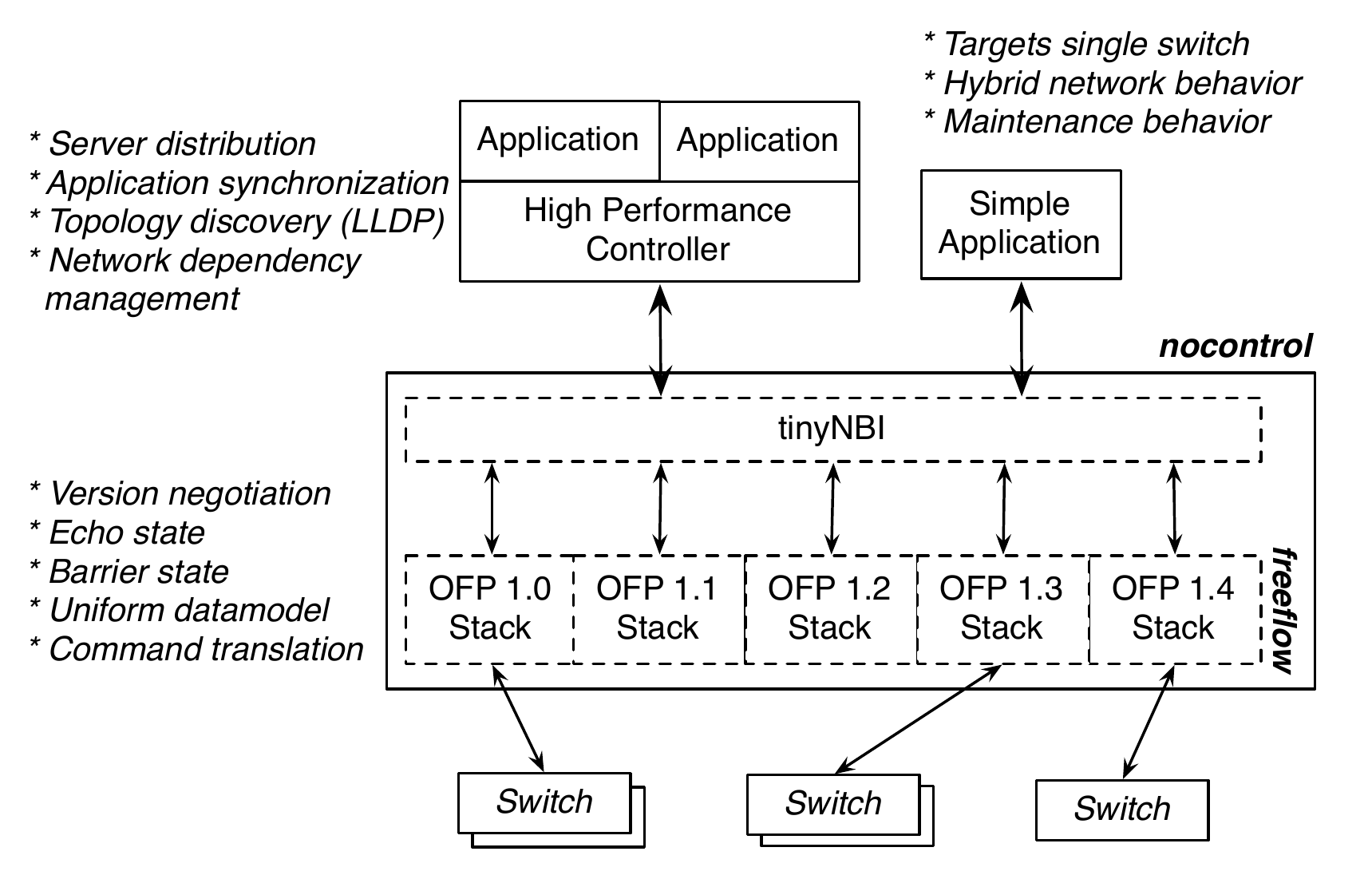}
    \caption{tinyNBI Relationships}
    \label{figure:topo}
  \end{center}
\end{figure}

Figure~\ref{figure:topo} shows the relationship of tinyNBI relative to an
OpenFlow protocol stack, a high performance controller, and OpenFlow
applications. We are in the process of implementing tinyNBI as part of the 
nocontrol controller \cite{freeflow} in order to support these applications.

%

\section{Background}

OpenFlow has a unique personality like no other network protocol. We were 
introduced to these peculiarities while building a low-level OpenFlow protocol
stack that supported versions: 1.0, 1.1, 1.2, 1.3.0, and 1.3.1. The stack 
provided facilities to create and manage: messages, state machines, system 
interfaces, and configurations. The intended use of this stack was to build 
controllers, switch-agents, applications, and benchmarks for experimentation. 
This work is now part of the Flowgrammable OpenFlow protocol
stack~\cite{freeflow}, 
and was a finalist in the ONF's OpenFlow Driver competition. With that
said, we were unsuccessful at keeping the interface simple in the face of
multiple versions of OpenFlow, and differing capabilities of switches. What
follows is a short description of the issues we encountered.

OpenFlow is a collection of standards, not a single standard. What most
people think of as OpenFlow is known as the core switch specification; however,
there are many other OpenFlow standards such as: configuration, test, and 
conformance. Within the core switch specification there are 
currently five published 1.X standards, several minor patches, and many
extension packages. There has been a new version of the standard every year for
the last five years, and unfortunately this trend does not seem likely to end 
soon. 

There is a high degree of variability between each version of the standard. Each
version of OpenFlow specifies an interface, and the collection of abstractions
present in a switch that can be manipulated. The types of abstractions and their
scope grows with each successive version of OpenFlow. Many of these 
abstractions are not mandated the standard. A properly behaving
application must first determine that a desired feature is provided under the
version of OpenFlow that was negotiated for the target switch, and then
determine if the target switch actually supports this feature. This process can
quickly turn even the simplest of applications into a mess with extremely
complex branching to test for all acceptable scenarios.

\begin{figure}[h]
  \begin{center}
    \includegraphics[width=.5\textwidth]{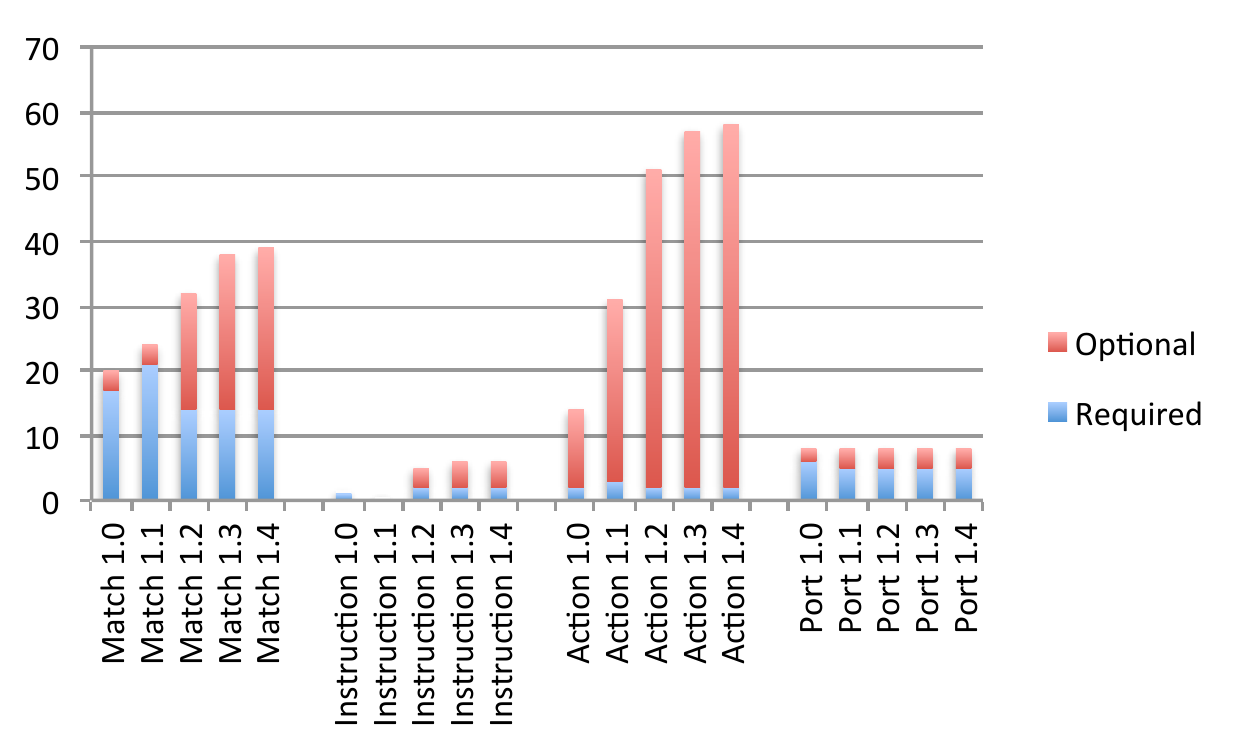}
    \caption{Variability in Capabilities}
    \label{figure:capabilities}
  \end{center}
\end{figure}

Figure~\ref{figure:capabilities} illustrates a of subset of abstractions from 
five versions of OpenFlow. Capabilities that switch must provide are shown as
`required', while capabilities defined but not mandated are shown as `optional'.
The general trend of new abstractions in each version is obvious. However, its 
more interesting to note that more than half of these abstractions are optional.
To make matters even more complicated, these capabilities may be hierarchical. 
Some of these abstractions come in sets, where each item of the set may have its
own unique capabilities.

The growth of abstractions in successive versions of the protocol have largely 
been additive. This is a useful property when considering a version agnostic
view of the abstractions. Unfortunately, the interface provided that manipulates
these abstractions has changed drastically. The method to use in
determining capabilities can change from each version of OpenFlow. For instance,
determining action capabilities is performed in three unique ways depending on 
version.

There are several programming languages that have been developed to model 
OpenFlow primitives. Some of these languages include: 
Frenetic~\cite{frenetic, netcore, netkat}, and Maple~\cite{nettle,
maple}. Our work differs from these languages in several ways. First, while
tinyNBI is specified in C, it is a language independent data model and
interface. We do not force users to use a specific language. Second, our
abstractions provide a complete set of OpenFlow semantics (meters, groups,
multiple flow tables, metadata, write, apply, actions, etc.). Third, our
contribution handles multiple concurrent OpenFlow versions and varying switch
capabilities without placing additional effort on the application developer.

Additionally, there has been some preliminary work to address the non uniformity
of capability support across target switches. NOSIX~\cite{nosix} presents a 
high-level interface and defines a mechanism for meeting low-level details of a 
target switch with driver development.

\section{OpenFlow Distilled}
\label{sec:model}

TinyNBI is a data model and interface that abstracts many the eccentricities of
OpenFlow, while maintaining a low-level of abstraction. The purpose of tinyNBI
is to provide a foundational interface for the development of higher level
network abstractions.

\begin{figure}[h]
  \begin{center}
    \includegraphics[width=.5\textwidth]{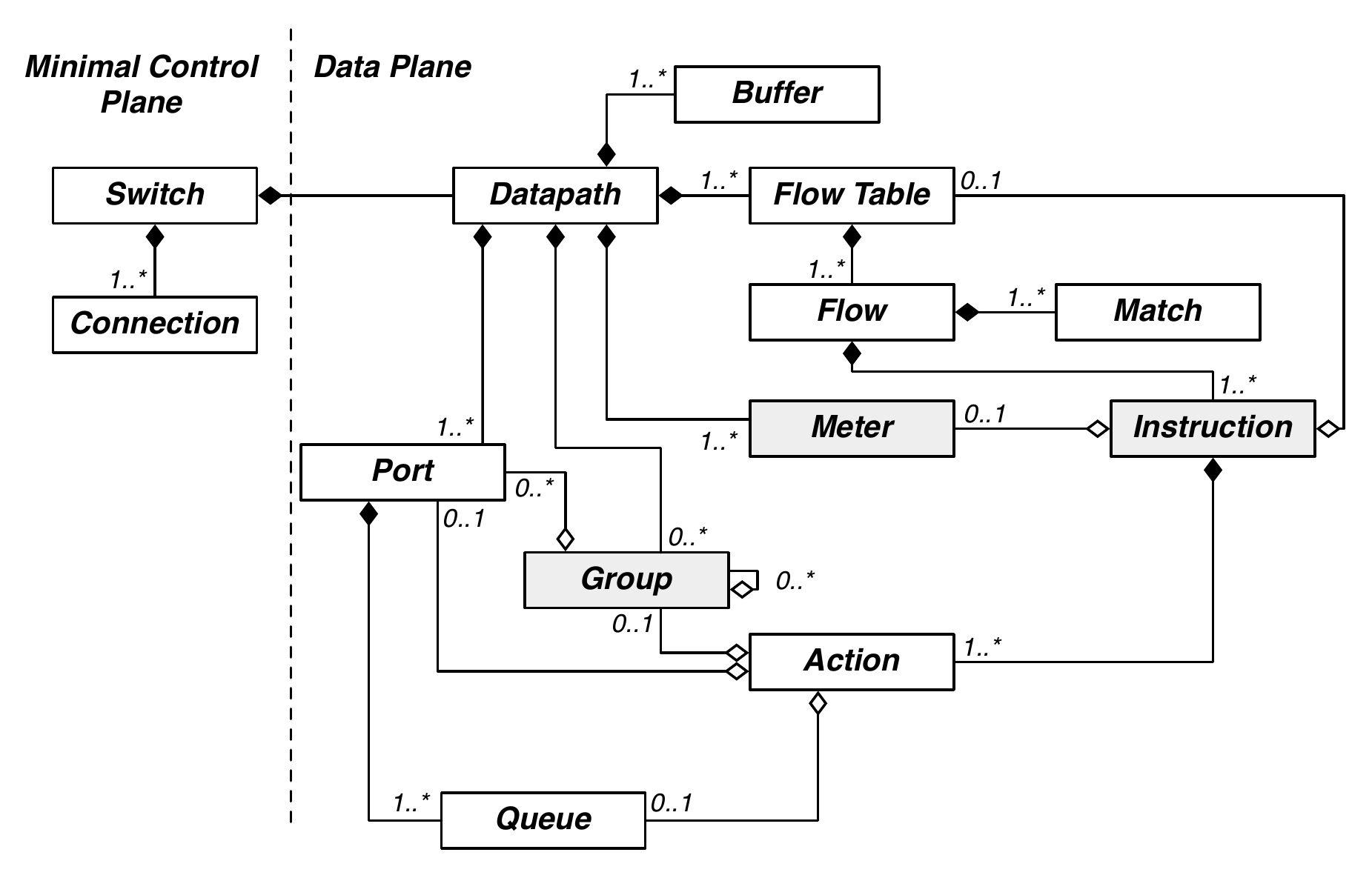}
    \caption{Unified Switch Datamodel}
    \label{figure:ofp_datamodel}
  \end{center}
\end{figure}

The core of tinyNBI is the data model, displayed in 
Figure~\ref{figure:ofp_datamodel}.
This data model makes a clear distinction between control plane and data plane
abstractions, both of which are necessary. Each abstraction has three
components: \textbf{configuration}, \textbf{capabilities}, and 
\textbf{statistics}. Configuration is data that is modifiable by the interface 
and changes the behavior of the abstraction. Capabilities are a non-modifiable 
state that describes the range of behaviors of an abstraction, which could be 
limited to a single behavior. Finally, statistics are read-only data that 
provides some description of how the abstraction has behaved.

Each abstraction present in Figure~\ref{figure:ofp_datamodel} can be
targeted directly with the tinyNBI API. OpenFlow itself sometimes requires
indirect methods to retrieve information pertaining to a specific abstraction.
The implementation of tinyNBI will ensure the appropriate OpenFlow interface is
used. Datapath is not an abstraction present in the OpenFlow specification; 
however, we feel it is a necessary root abstraction for the data plane side of a
switch. It is more meaningful to a developer to ask the table capacity and
fragmentation reassembly behavior of the datapth.

Another important observation is that four of the abstractions represent finite
addressable resources: flow tables, flows, meters, and groups. Programmatically,
resources are notoriously prone to leakage (memory, sockets, etc.). While 
OpenFlow combines the act of allocation and initialization TinyNBI
provides programmers with a clear separation between these activities. This is 
to allow familiar resource management techniques, such as Resource 
Initialization as Acquisition (RAII), to be used for safer program behavior.

Not all abstractions are present in all versions of OpenFlow, for instance the
gray boxes in Figure~\ref{figure:ofp_datamodel} are not present in version 1.0 
of standard. Abstractions absent in a switch are handled in one of three ways: 
seamless emulation, switch offload, or error indication. Seamless emulation
involves defining the abstraction but with a limited set of capabilities. These
capabilities are then provided by using semantically equivalent operations that 
are present. For instance, 1.0 action sets are equivalent to the 1.1$^+$ Instruction
Apply. Switch offload involves offloading the missing switch functionality to
the controller. While this procedure will not be as efficient as switch 
processing, it can give the appearance of seamless functionality. 

As has been demonstrated in both the security and graphics communities, having a
common interface with software offload when hardware is not present can still be
quite useful. Finally, there are some scenarios where it is never acceptable to
provide the missing behavior. In these scenarios we provide the calling
application with an explicit error indication.

\section{Programming Interface}
\label{sec:api}

tinyNBI is not a high level interface; it only provides access to the elements
of the abstract model, their properties, and major functionality. It does not
try, for example, to codify support for representing network topology,
switching, routing, load balancing, provisioning, or even operational aspects of
network management. We see those as applications or abstractions releying on
a solid foundation: the tinyNBI model. The API provides direct support for
the creation of higher level applications and programming models.

A particular concern of the API is independence from the ever-changing structure
and semantics of the OpenFlow specification. The abstract switch described in
Section~\ref{sec:model} provides a basis for an API that evolves  incrementally.
That is, the addition of new features does not modify the structure or semantics
of existing model elements. The API also embraces this idea by hiding data
and functionality behind a procedural interfaces. The amount of data,
represented by structures or records, is minimal.

The API is specified as an interface in the C Programming Language. We choose
this approach in order to achieve maximum portability achieve greater re-use
potential, and is inspired by the POSIX specification and typical system call 
interface of C runtime libraries. 

Elements of the model are accessed using descriptors: integer values that
denote an element in the model. Direct access to those structures is not
available. Arguments passed to the procedural interface are passed as
integers or pointers to (mostly) opaque types. This allows the tinyNBI
implementation to change over time without breaking user applications or
requiring re-compiles. 

Most of the procedures in the API are variadic functions whose behavior is 
determined by a \emph{selector}: an integer value selecting an abstraction
to query or modify or a behavior to invoke. This approach means that 
applications written against older versions of the library will be binary
compatible with the newer versions. Requiring recompilation of applications
for each new version of the protocol will drive users away.  A partial listing
of element selectors and those operations is shown below:
\begin{program}
enum ofp_selector {
  OFP_CONTROLLER, OFP_SWITCH, OFP_CONNECTION, OFP_PORT, 
  OFP_QUEUE, OFP_TABLE, OFP_FLOW, OFP_GROUPS, OFP_GROUP, 
  OFP_METERS, OFP_METER, OFP_MATCH, OFP_INSTRUCTION, 
  OFP_ACTION, OFP_ERROR, OFP_EXTENSION, ...
};
\end{program}
Here, each macro denotes a primary abstraction in the model with the
exception of \code{OFP_CONTROLLER} and \code{OFP_SWITCHES}. The former
provides access to the controller itself, and the latter applies to
the set of connected switches. Every other value refers to specific
elements in the model. The \code{OFP_GROUPS} and \code{OFP_METERS}
selectors describe the group table and meter table.

The API is comprised of only a handful for commands. We present them as 
collections of related functionality: application
requirements, resource acquisition, queries and commands, event processing, and
application lifecycle, and extensions. Most commands are defined in
terms of a connected switch. Connected switches can be queried from the
controller (\secref{sec:api.life}) or notified as a signal
(\secref{sec:api.event}).

\subsection{Program Lifecycle}
\label{sec:api.life}

An OpenFlow application is written against an OpenFlow controller using the
tinyNBI API. The API supports two ways in which the application and
the controller interoperate:
\begin{itemize}
\item The application drives the controller. In this case, the
application is responsible for its own setup and causing the controller
to process the OpenFlow protocol.
\item The controller Drives the application. Effectively, the application
is a module or plugin responding asynchronously to controller events.
\end{itemize}
tinyNBI accommodates both views. See Section~\ref{sec:api.event} for a
description of synchronous and asynchronous event handling. An application
truly begins when it is applied to a switch. Discovery of switches can happen in
one of two ways: the controller can be queried for a set of connected switches
or can be notified of a new switch. Below is a sample program that demonstrates
querying for all connected switches:
\begin{program}
struct ofp_switch_id* sws;
int len;
int err = ofp_get(OFP_CONTROLLER, OFP_SWITCHES, 
                  OFP_SWITCH_ALL, &sws, &len);
if (erro < 0)
  ofp_perror("error");
\end{program}
The \code{ofp_get} function issues queries against an abstraction indicated
by a selector. The arguments following the selector depend on the value
provided. This call allocates an array of \code{len} switch ids, each
corresponding to a connected switch. An error occurs if the result is negative.

Once a connected switch has been identified, the application has
the following lifecycle.
\begin{enumerate}
\item Declaration -- Announces program requirements
\item Construction -- Acquires necessary resources
\item Execution -- Creates, modifies, removes flows
\item Destruction -- Releases acquired resources
\end{enumerate}
These phases approximate the lifetime of an object in the C++ programming
language. We describe each phase separately in the following subsections.

\subsubsection{Requirement declaration}
\label{sec:api.req}
Prior to execution, a program must announce its requirements. This is done
through a sequence of declarations using the \code{ofp_require} command.
For example, a simple learning bridge application requires two flow tables
and the ability to match ingress ports and Ethernet source and destination
addresses. Here, \code{s} denotes a connected switch and \code{off} is an
\emph{offload} flag (described below).
\begin{program}
bool check_requirements(ofp_switch_id s, bool off) {
  return 
  ofp_require(OFP_SWITCH, s, off, OFP_TABLES, 2)    &&
  ofp_require(OFP_MATCH, s, off, OFP_FIELD_IN_PORT) &&
  ofp_require(OFP_MATCH, s, off, OFP_FIELD_ETH_SRC) &&
  ofp_require(OFP_MATCH, s, off, OFP_FIELD_ETH_DST);
}
\end{program}
If any call is \code{false}, then the application cannot be executed on the
connected switch. This approach to declaring requirements before execution
allows programmers to avoid repeated checks for specific capabilities of
connected switch\-es. Note that this does not guarantee program correctness.
Operations may still fail if, for example,the switch runs out of memory for new
flows.

Note that requiring two tables only checks that the switch contains at least
two tables. It does not require that two tables are currently available. 
Flow tables must be explicitly requested during resource acquisition 
(\secref{sec:api.raii}).

The \emph{offload} flag is used to indicate that tinyNBI should partially 
offload packet processing to the controller if the connected switch does
not fully support the required operation. For example, if \code{s}
is not (for some reason) capable of matching Ethernet fields, then tinyNBI
can install a flow that redirects the packet to the controller where
processing is done in software.

Feature offloading helps provide a consistent platform for OpenFlow application
programmers. However, not all features can be effectively offloaded and the
process may not be viable for high-performance controllers or applications.
In that case, the offload flags should be \code{false} and, which means the
\code{ofp\_requires} will fail if the feature is not provided by the switch.

\subsubsection{Resource Acquisition}
\label{sec:api.raii}

Having announced requirements, the application must acquire its necessary
resources. For example our learning bridge requires two tables: one to learn
Ethernet source and addresses, and the other for forwarding.  This is done using
the \code{ofp_acquire} command. Resources are released during destruction with
the \code{ofp_release} command. This model of is precisely the RAII (Resource
Acquisition Is Initialization) idiom used in C++.
Acquiring a table for the learning bridge application can be written this
way.
\begin{program}
struct ofp_flow_table learn;
if (ofp_acquire(OFP_TABLE, s, &learn) < 0) {
  ofp_perror("cannot acquire learning table");
  exit(-1);
}
\end{program}

Failing to acquire this table means the program cannot continue running. 
Allowing an OpenFlow application to continue running in an uninitialized or
partially initialized state invites opportunities for errors or vulnerabilities.

When the application has completed or been terminated by an operator, it must
release any acquired resources.
\begin{program}
ofp_release(&learn);
\end{program}

Note that groups and meters are also considered as resources in this model.
An OpenFlow application that uses either must explicitly acquire (and release)
those resources prior to execution.

\subsubsection{Event handling}
\label{sec:api.event}
An application responds to events in one of two ways: synchronously or
asynchronously. At the heart of the event system is the \code{ofp_event}
type, which encodes information about specific events as a \code{union}
of event structures.

Synchronous event handling is done using the \code{ofp_poll}
function. This is ideal when the OpenFlow application is driving the
controller. For example:
\begin{program}
struct ofp_event e;
while (!ofp_poll(10, &e)) {
  switch (e.type) {
  case OFP_SWITCH_EVENT:
    on_switch_event(e.switch_id, e);
    break;
  case OFP_PACKET_EVENT:
    on_packet_event(e.switch_id, e);
    break;
  }
}
\end{program}
The \code{ofp_event} type encodes the event type and originating switch
(as \code{switch_id}). A switch event can include the include the connection
or disconnection of a switch. This allows the application to free resources
or terminate other applications. The packet event is essentially an OpenFlow
packet-in event.

Alternatively, an application may register handlers for specific
events. This is useful when the controller is driving, and the application
is a module or plugin. For example:
\begin{program}
// Event handlers
int on_switch_event(ofp_switch_id, const ofp_event*);
int on_packet_event(ofp_switch_id, const ofp_event*);

// Register handlers
ofp_register(on_switch_event, OFP_SWITCH_EVENT);
ofp_register(on_packet_event, OFP_PACKET_EVENT);
\end{program}
Event selectors are a bitfield, so a single handler may respond to multiple
events. 

\subsection{Queries and Commands}
\label{sec:api.query}
Queries and commands are used to request information about a dataplane
primitive or a set of such primitives. There are three primary operations
related to these abstractions:
\begin{itemize}
  \item \code{ofp_get} --- Get configuration properties
  \item \code{ofp_set} --- Modify configuration properties
  \item \code{ofp_stats} --- Get updated statistics
\end{itemize}
For example, getting properties about a port and its statistics can be 
done using a sequence of commands:
\begin{program}
struct ofp_port p;
struct ofp_port_stats ps;
ofp_get(OFP_PORT, s, OFP_PORT_CONTROLLER, &p);
ofp_stats(OFP_PORT, s, OFP_PORT_CONTROLLER, &ps);

printf("%s", p.addr);
printf("%d\n", p.rx_packets);
\end{program}
Here, we assume the operations succeed, however the results should be checked
to avoid invoking undefined behavior.

Updating a port is done using the \code{ofp_set} operation. For example,
an application can try shutting down all ports.
\begin{program}
struct ofp_port pm;
pm.port_id = OFP_PORT_ALL;
pm.config = OFP_PORT_DOWN;
if (ofp_set(OFP_PORT, s, OFP_PORT_CONTROLLER))
  exit(-1);
\end{program}

The interface provides two additional operations for working with flows.
These are thin wrappers around \code{ofp_set} that provide some default
arguments for modifying a flow entry.
\begin{itemize}
  \item \code{ofp_add} --- Add a flow to a flow table.
  \item \code{ofp_del} --- Remove a flow from a flow table.
\end{itemize}
For example, adding a new flow is done using the following program.
\begin{program}
struct ofp_flow learn_all {
  .table = &learn, /* learning table */
  .match = /* unspecified */,
  .exec = /* unspecified */
};
if (ofp_add(s, OFP_FLOW, &learn_all) < 0)
  syslog(LOG_ERROR, "oh no!");
\end{program}
The \code{ofp_flow} type encodes the match and instructions being added
to the switch. The \code{match} and \code{exec} fields specify the matching
condition and corresponding instruction set to be executed. These are described
in \secref{sec:api.flow}.

Removing a flow is similar. For example, in a learning
bridge, we want to remove forwarding flows whenever a learned flow times
out. The command to explicitly remove that flow is:
\begin{program}
struct ofp_flow forget_fwd {
  .table = &forward, /* forwarding table */
  .match = /* unspecified */
}
if (ofp_del(s, OFP_FLOW, &forget_fwd) < 0)
  syslog(LOG_ERROR, "oh no!");
\end{program}

\subsection{Flow construction}
\label{sec:api.flow}
Flow construction is achieved through a switch-independent interface comprised
of three functions:
\begin{itemize}
\item \code{ofp_build_match}
\item \code{ofp_build_action}
\item \code{ofp_build_intstrution}
\end{itemize}
Matches, actions, and instructions are built incrementally through a sequence
of the calls above. The \code{ofp_flow} structure above binds matches and
instructions when adding, modifying, and deleting flows. For example, the
initial match for a learning table matches all inbound packets, outputs them
to the application, and then sends to the forwarding table. There are
three sets of operations involved:
\begin{program}
// Construct a match
struct ofp_match all;
ofp_make_match(&all, NULL); // Matches everything

// Build the default learning instruction.
struct ofp_actions out;
ofp_build_action(&out, OFP_OUTPUT, OFP_PORT_CONTROLLER);

// Construct the instruction set.
struct ofp_instructions ins;
ofp_build_instruction(&ins, OFP_APPLY, &out);
ofp_build_instruction(&ins, OFP_GOTO, &forward);

struct ofp_flow learn_all {
  .table = &learn, /* learning table */
  .match = &all,   /* match all packets */
  .exec = &ins     /* execute instructions */
}
\end{program}
Note that matched fields, actions, and instructions used must be declared as
requirements prior to application execution. The flow \code{learn\_all} can be
installed on a switch as needed.

Consider another example where the application, when responding to a packet
in event, constructs a new flow that specifically matches the port and
Ethernet source address. The program first extracts that information from a
packet object in an \code{ofp_event} structure and builds a flow to match it:

\begin{program}
struct ofp_port_id port;
ofp_get(OFP_PACKET, OFP_FIELD_IN_PORT, e.packet, &port);

char src[6];
ofp_get(OFP_PACKET, OFP_FIELD_ETH_SRC, e.packet, src);

struct ofp_match match;
ofp_build_match(&match, OFP_FIELD_IN_PORT, port);
ofp_build_match(&match, OFP_FIELD_ETH_SRC, src);
\end{program}
This match will be added as a flow in the \code{learn} table.

\subsection{Extensions}
\label{sec:api.extensions}
Experimenter (or vendor) extensions are an important aspect of the
OpenFlow ecosystem. Our current approach to providing extension features
builds on the same model for working with other abstractions. We represent
all experimenter extensions as a binary block of data and its size. No
validation or checking is done on the contents of that data.

The \code{OFP_EXTENSION} selector is accepted by most operations in the 
API, and is typically followed by three or four arguments, depending on the
operation. The \code{ofp_get} and \code{ofp_set} operations takes the 
experimenter (or vendor) id and the type of message. The last two arguments
are always a pointer to the experimenter data and its size.
Both \code{ofp_get} and \code{ofp_set} send an OpenFlow vendor or experimenter 
message. Currently, \code{ofp\_get} expects to 
the extension data to be an in-out parameter, while \code{ofp\_set} takes
the data to be constant. 

\begin{program}
struct sw_info inf {
  char major, minor, patch;
};
if (!ofp_get(OFP_EXTENSION, eid, SW_INFO, inf, n))
  printf("%d.%d.%d", inf.major, inf.minor, inf.patch);
\end{program}

The functions \code{ofp_stats} also accepts an \code{OFP_EXTENSION}, and sends a
stats request with the appropriate codes. The extension data is populated
directly from the contents of the return message. The flow constructors for
matches, actions, and instructions behave similarly. The extension data is
copied directly into the corresponding message structures and sent as part of
the enclosing message.

\section{Conclusion and Future Work}

Writing OpenFlow applications that operate correctly in robust network
environments is a tough proposition. Building high level network abstractions
over this shifting landscape seems like a difficult proposition. We propose
tinyNBI as a low-level NBI to provide a stable foundation for designing these
high level abstractions. We are currently designing higher-level abstractions on
top of tinyNBI that allow for clean separation of operator versus developer
concerns.

\bibliographystyle{plain}
\bibliography{references}

\end{document}